\documentclass[aps,prc,twocolumn,showpacs,preprintnumbers,superscriptaddress,
amsmath,amssymb]{revtex4}
\usepackage{graphicx}
\usepackage{dcolumn}
\usepackage{bm}

\begin{document}
\def\GSI{Gesellschaft f{\"u}r Schwerionenforschung mbH, D-64291 Darmstadt,
Germany}
\def\MOSCOW{Institute for Nuclear Research, 117312 Moscow, Russia}
\def\KONYA{Department of Physics, University of Sel\c{c}uk, 42079 Konya, 
Turkey}
\def\IFJ{H. Niewodnicza{\'n}ski Institute of Nuclear Physics, 
Pl-31342 Krak{\'o}w,
Poland}
\def\FIAS{Frankfurt Institute for Advanced Studies, J.W. Goethe University, 
D-60438 Frankfurt am Main, Germany}
\def\KURCH{Kurchatov Institute, Russian Research Center, 123182 Moscow, Russia}
\newcommand{\goo}{\,\raisebox{-.5ex}{$\stackrel{>}{\scriptstyle\sim}$}\,}
\newcommand{\loo}{\,\raisebox{-.5ex}{$\stackrel{<}{\scriptstyle\sim}$}\,}

\title{Modification of surface energy in nuclear multifragmentation.}

\affiliation{\GSI}
\affiliation{\MOSCOW}
\affiliation{\KONYA}
\affiliation{\IFJ}
\affiliation{\FIAS}
\affiliation{\KURCH}

\author{A.S.~Botvina}       \affiliation{\GSI}\affiliation{\MOSCOW}
\author{N.~Buyukcizmeci}    \affiliation{\KONYA}
\author{M.~Erdogan}     \affiliation{\KONYA}
\author{J.~{\L}ukasik}          \affiliation{\GSI}\affiliation{\IFJ}
\author{I.N.~Mishustin}     \affiliation{\FIAS}\affiliation{\KURCH}
\author{R.~Ogul}        \affiliation{\GSI}\affiliation{\KONYA}
\author{W.~Trautmann}           \affiliation{\GSI}

\date{\today}

\begin{abstract}
Within the statistical multifragmentation model we study modifications 
of the surface and symmetry energy of primary fragments in the freeze-out
volume. The ALADIN experimental data on multifragmentation obtained in
reactions induced by high-energy projectiles with different
neutron richness are analyzed. We have extracted the isospin dependence of 
the surface energy coefficient at different degrees of fragmentation. 
We conclude that the surface energy of hot fragments 
produced in multifragmentation reactions 
differs from the values extracted for isolated nuclei at low excitation.
At high fragment multiplicity, it becomes nearly 
independent of the neutron content of the fragments.
\end{abstract}


\maketitle

\section{Introduction}
\label{intro}

A break-up of nuclei into many fragments (multifragmentation) has
been observed in nearly all types of nuclear reactions when a large amount
of energy is deposited in nuclei. It includes reactions
induced by protons, pions, antiprotons, and by heavy ions of both, 
relativistic energies (peripheral collisions) and 
'Fermi'-energies (central collisions). According to the present
understanding, multifragmentation is a relatively fast process,
with a characteristic time around 100 fm/c, where, nevertheless, a high
degree of equilibration is reached. The process is mainly
associated with abundant production of intermediate mass fragments
(IMFs, with charges $Z \approx$ 3-20). However, at the onset of
multifragmentation, also heavy residues are produced which have  
previously only been associated with compound-nucleus processes. At
very high excitation energies, the IMF production gives way
to the total vaporization of nuclei into nucleons and very light clusters.

The previous ALADIN experiments have provided extensive
information about multifragmentation of projectiles in peripheral
nucleus-nucleus collisions at high energy \cite{ALADIN}. They have
demonstrated a 'rise-and-fall' of multifragmentation with
excitation energy, and they have shown that the temperature remains nearly
constant, around $T\sim$5 MeV, during this process
\cite{Pochodzalla}. There are large fluctuations
of the fragment multiplicity and of the size of the largest fragment in the
transition region from a compound-like decay to full
multifragmentation of spectators \cite{kreutz}.
It was found that the statistical models, which assume a thermal 
equilibration among hot fragments in a freeze-out volume at subnuclear 
densities, are fully consistent with the data 
\cite{Botvina92,SMM,Gross,Raduta}. 

We believe that multifragmentation studies pursue two main
purposes. The first one is connected with the general understanding
and better description of this reaction which represents as much
as 10-15$\%$ of the total cross section in high energy
hadron-nucleus collisions, and nearly twice more in
nucleus-nucleus collisions. Secondly, the multifragmentation
reaction can be considered as an experimental tool to study the
properties of hot fragments and the phase diagram of nuclear
matter at densities $\rho \approx 0.1-0.3 \rho_0$ and temperatures around $T
\approx$ 3--8 MeV which are expected to be reached in the
freeze-out volume ($\rho_0\approx0.15$~fm$^{-3}$ is the
normal nuclear density). Multifragmentation opens a unique possibility to
investigate this part of the phase diagram and to determine the 
"in-medium" modifications of nuclei there. This second point is very
important for many astrophysical applications, in particular, for
processes during the supernova II explosions and neutron star
formation \cite{Bethe,Botvina04,Botvina05}.

Some of the properties of hot fragments have been addressed in our
previous work. For example, the symmetry energy was extracted in
ref. \cite{LeFevre}, and it was demonstrated that it decreases 
considerably with excitation energy. In this paper we show that also the
surface energy of hot fragments can be investigated in these
reactions, and that the modified surface energy can be extracted by comparing 
theory with experiment.

\vspace{0.5cm}

\section{Statistical approach for the description of multifragmentation}

Statistical models are used in situations when an equilibrated
source can be defined in the nuclear reaction. The most famous
example of such a source is the 'compound nucleus' introduced by
Niels Bohr in 1936. The standard compound nucleus picture is valid
only at low excitation energies when sequential evaporation of light
particles and fission are the dominating decay channels. However,
this concept cannot be directly applied at high excitation energies, $E^*
\goo $ 2--3 MeV/nucleon, when the nucleus rapidly
disintegrates into many fragments. As was shown in many
experiments (see e.g. \cite{Botvina95,EOS,ISIS,MSU,INDRA,FAZA}),
an equilibrated source can be formed in this case too, and
statistical models are generally very successful in describing the 
fragment production.

As a basis for our study below we take the Statistical
Multifragmentation Model (SMM), for a review see ref. \cite{SMM}. The model
assumes nuclear statistical equilibrium at a low-density
freeze-out stage. It considers all break-up channels composed of
nucleons and excited fragments taking into account the
conservation of baryon number, electric charge and total energy. Light 
clusters with mass number A $\leq 4$ are treated as elementary
particles with only translational degrees of freedom ("nuclear
gas"). Fragments with A $> 4$ are treated as heated liquid drops. In
this way one may study the nuclear liquid-gas coexistence in the
freeze-out volume. The Coulomb interaction of fragments is
described within the Wigner-Seitz approximation. Different decay 
channels $f$ are generated by Monte Carlo sampling according to
their statistical weights, $\propto \exp{S_f}$, where $S_f$ is the
entropy of channel $f$. After the break-up, the Coulomb
acceleration and the secondary de-excitation of primary hot
fragments are taken into account. The SMM has been already successfully 
applied for the analysis of ALADIN data \cite{Botvina95,Xi}.

\section{Influence of the symmetry energy on multifragmentation}
\label{symm}

In the SMM, the symmetry energy of hot fragments with mass
number $A$ and charge $Z$ is parametrized as
$E^{\rm sym}_{A,Z}=\gamma (A-2Z)^2/A$, where $\gamma$ is a
phenomenological parameter. In the case of cold nuclei $\gamma
\approx 25$~MeV is adopted in order to describe their 
binding energies. For hot fragments this parameter can be modified and, 
therefore, should be extracted from experimental data. The corresponding 
set of isospin-related data may be provided by the multifragmentation 
reactions, e.g., via the isoscaling phenomenon \cite{Botvina02}.
At present, there are evidences for a significant reduction of the
symmetry energy in hot nuclei \cite{LeFevre,Shetty05}.

\begin{figure}

\includegraphics[width=9cm]{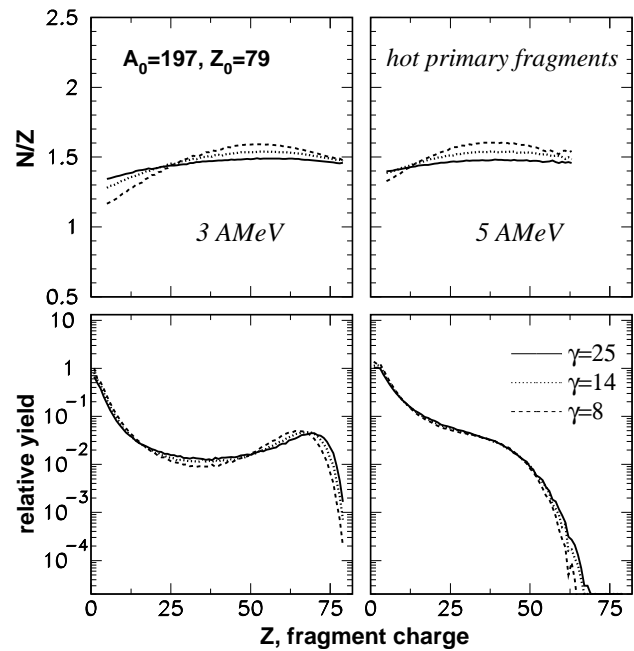}
\caption{\small{Influence of the symmetry energy coefficient
$\gamma$ on yields of hot fragments in the freeze-out volume
(bottom panels) and on their $N/Z$ ratios (top panels), in SMM
calculations for Au sources at excitation energies of 3 and 5
MeV/nucleon (left and right panels, respectively). }}
\end{figure}

The influence of the symmetry energy on the characteristics of produced 
fragments was already partly analyzed in \cite{nihal}.
It was shown that it has a small effect on average quantities such as 
the temperature and fragment masses, but it 
influences mainly the isotope distributions of the fragments. 
This is illustrated in Fig.~1 for the mean charge distributions and
neutron-to-proton ($N/Z$) ratios of fragments.
These characteristics change only slightly even with rather large
variations of $\gamma$. In this respect, we believe that variations of 
$\gamma$ will 
not influence the results of previous analyses of experimental
data related to the fragment charge partitions and average isotope yields
\cite{Botvina95,EOS,ISIS,MSU,INDRA,FAZA,SMM,Xi}.

\begin{figure}
\includegraphics[width=8.5cm]{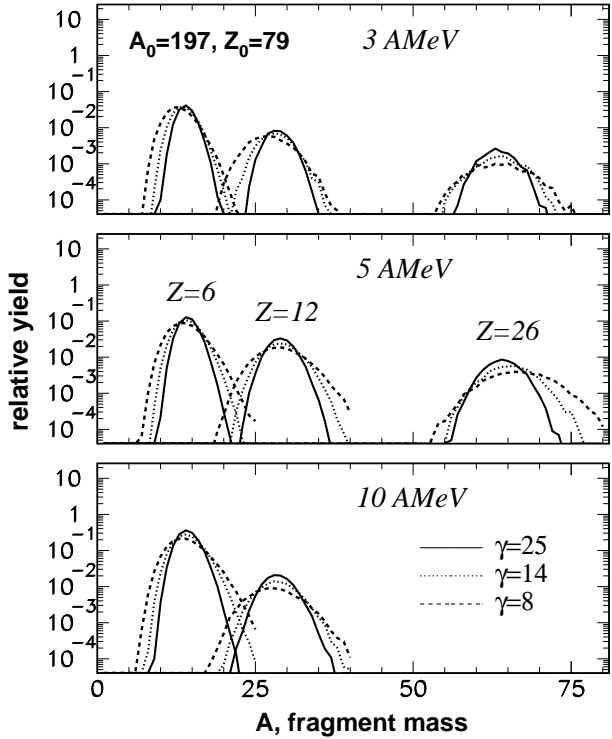}
\caption{\small{Mass distributions of hot fragments with atomic numbers
$Z=$ 6, 12, and 26, produced in the multifragmentation of Au sources
at excitation energies of 3, 5 and 10 MeV/nucleon, for different
symmetry energy coefficients $\gamma$. }}
\end{figure}

By contrast, the symmetry energy has a large effect on isotope distributions. 
As evident from Fig.~2, they become wider at small $\gamma$
\cite{Botvina02,Botvina01}. The subsequent deexcitation of the produced hot
fragments changes the distributions in two ways: their widths usually 
become narrower and their centroids are shifted toward symmetry 
\cite{nihal,Botvina02,Botvina87}. This effect has to be 
taken into account when comparing with experimental data; it is essential, 
e.g., for the isoscaling analysis \cite{LeFevre,Botvina02,Shetty05}.

In the following, we study possible modifications of the fragment 
surface energy employing different isospin compositions of the sources.
Within the SMM this issue can be separated from modifications of the 
symmetry energy because the latter does not significantly change the mean 
charge yields. 
In phenomenological mass formulae, isospin dependent terms for the surface 
energy are sometimes introduced in order to obtain a better description of 
the ground-state masses of nuclei \cite{Cameron,Myers}. However, some modern 
mass formulae can provide even better descriptions of the nuclear masses 
retaining only the bulk isospin 
term but with a special treatment of the shell effects \cite{Nayak}. 
We believe that the best strategy for the present purpose of 
investigating new 'in-medium' properties of nuclear fragments is to 
consider a minimum number of parameters which have a clear physical meaning. 
We, therefore, use the standard SMM liquid-drop parametrization which 
contains no surface isospin term. Instead we allow for the surface 
coefficient to be a free parameter which is determined by comparing 
the calculations with  
experimental data obtained for multifragmentation of nuclei with different 
isospin content.

\section{Influence of the surface energy on multifragmentation}
\label{surf}

The surface free energy of hot fragments in the SMM is
parametrized as $F^{\rm surf}_{A,Z}=B(T)A^{2/3}$, where 
$B(T) = B_0 [(T_{\rm c}^2-T^2)/(T_{\rm c}^2+T^2)]^{5/4}$.
Here $B_0\approx$18 MeV is the surface coefficient for isolated
cold nuclei, as adopted in the Weizsaecker mass formula, and 
$T_{\rm c} \approx$
18 MeV is the critical temperature for the nuclear liquid-gas
phase transition in infinite matter. 
One should distinguish $T_{\rm c}$ from the phase transition temperature 
in finite hot nuclei, which is essentially lower, around 5--6 MeV 
\cite{SMM}. $T_{\rm c}$ should be considered as a model parameter 
characterizing the temperature dependence of the surface tension in 
finite nuclei. At low temperatures this $T$-dependence leads to a correct 
surface contribution to the level densities of nuclei, i.e., it describes 
correctly the heat capacities of isolated nuclei. The surface parameters 
may also be modified in the low-density nuclear medium, i.e., in 
environments consisting of nucleons and hot fragments. The possible 
changes of $T_{\rm c}$ in multifragmentation reactions were analyzed in
refs. \cite{ogul,karna,nihal}, and from the analysis of experimental 
data it was found that $T_{\rm c} \approx$~17--20 MeV in reactions 
with Au nuclei. The $T_{\rm c}$ effect on fragment yields is rather
small and, in any case, it is included in the surface energy. 
In the present work we concentrate on the influence of the $B_0$ coefficient 
on the multifragmentation pattern. 

The surface energy is rather important, since production of new
fragments means increasing the surface contribution to the total
energy of the system.
Therefore, even small variations of the surface energy lead to big
changes of fragment mass and charge distributions. This is demonstrated 
in Fig.~3 for
different excitation energies of Au sources with a freeze-out
density $\rho = \rho_0/3$. Decreasing the surface energy favors the
disintegration already at low excitation energies while
a larger $B_0$ suppresses the multifragmentation channels.

\begin{figure}
\includegraphics[width=9cm]{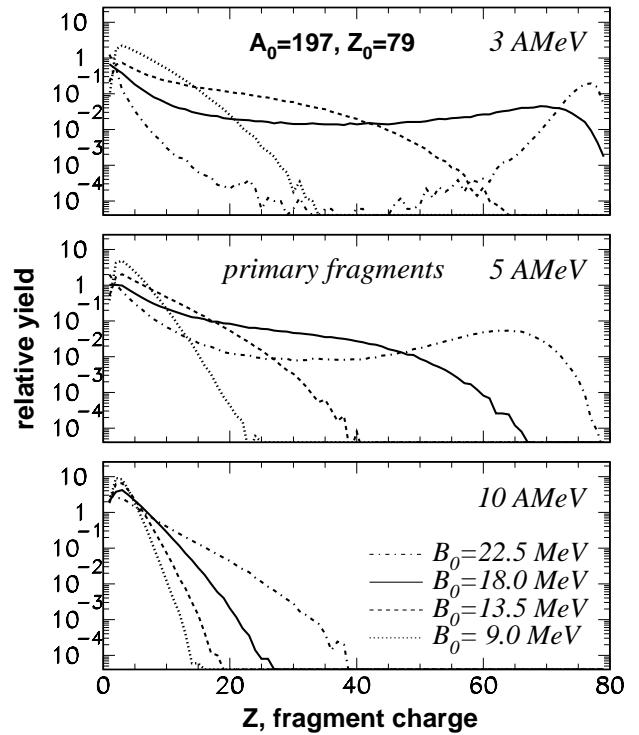}
\caption{\small{Influence of the surface energy coefficient $B_0$
on charge yields of hot fragments in the freeze-out volume, as obtained from
SMM calculations for Au sources at excitation energies of 3, 5, and
10 MeV/nucleon.}}
\end{figure}

To characterize the charge distributions we use the $Z^{-\tau}$ fit of the 
fragment yields. In order to avoid contributions of other processes
leading, e.g., to fission-like large fragments ($Z>20$) and to
coalescence-like small clusters ($Z \loo 2$), we limit the fragment 
charges by the range $Z=5-15$ in this fit. 
The extracted $\tau$ parameters for three sources with
different isospin, $^{238}$U, $^{197}$Au and $^{129}$Xe, are shown in Fig.~4.
In these calculations, the surface coefficient was fixed at the 
standard value of $B_0=18$~MeV. One can see that the isospin 
effect is rather essential, as was already discussed in
\cite{ogul,nihal}.

\begin{figure}
\includegraphics[width=8cm, height=7cm]{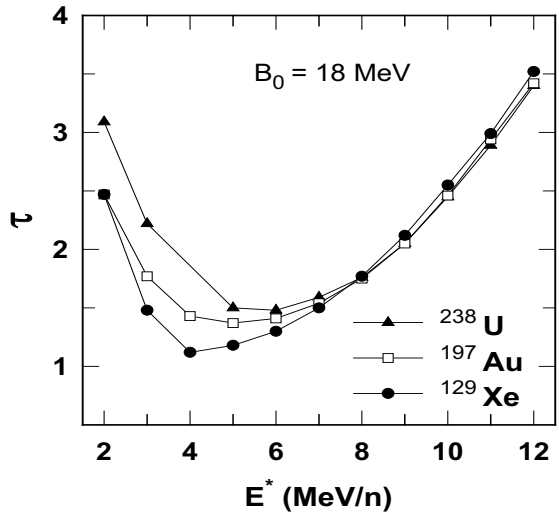}
\caption{\small{SMM calculations of $\tau$ parameters for hot
fragments from $^{238}$U, $^{197}$Au and $^{129}$Xe sources as a function 
of their
excitation energies. The power-law fitting was performed for
fragments with $Z=5-15$. }}
\end{figure}

We choose the Au source to demonstrate possible effects of 
the modified surface energy on multifragmentation. Figure~5 shows the
influence of the coefficient $B_0$ on the caloric curve, on the mass
number $A_{\rm max}$ of the largest fragment, and on $\tau$. We should
point out that thermodynamical characteristics of the system in
the freeze-out volume can also be influenced by the modifications of 
the surface 
energy of fragments. When these hot fragments leave the freeze-out
volume and decay, their normal properties are restored.
In order to emphasize the difference between isolated 
nuclei and nuclei in the medium, we introduce two different temperatures. 
One is an effective temperature, $T_{\rm eff}$, which is found from 
the energy balance in the freeze-out volume by assuming that the
properties of fragments are the same as those of 
isolated nuclei. This effective temperature 
reflects internal excitations of fragments respective to their
ground states and, thus, it can be compared to the temperature 
reached in the compound nucleus at the same excitation. 
Another temperature is the freeze-out temperature $T$
which includes in-medium modifications of the fragment properties, 
i.e., different $B_0$. The evolution of both temperatures with $E^{*}$ 
is presented in Fig.~5 in the
top two panels. One can see that variations in $B_0$ lead to noticeable 
effects. With a small $B_0$, correlated with a smaller effective temperature, 
the system disintegrates into lighter fragments, and 
$A_{\rm max}$ becomes smaller too. At low excitations the effective and 
freeze-out temperatures exhibit a similar behavior. However,
at very high excitation energies when the system disintegrates
only into light IMFs, the freeze-out temperature becomes higher at
smaller $B_0$. This behavior has a simple explanation:
the smaller surface term leads to a 
smaller level density parameter for internal excitation of fragments, which 
requires higher freeze-out temperatures in order to achieve the energy
balance. The $\tau$ parameter exhibits a general trend in the evolution 
of the charge 
distributions, it depends strongly on the excitation energy for all 
considered values of the surface coefficient (Fig.~5, lower panel). 

\begin{figure}
\includegraphics[width=8cm,height=14cm]{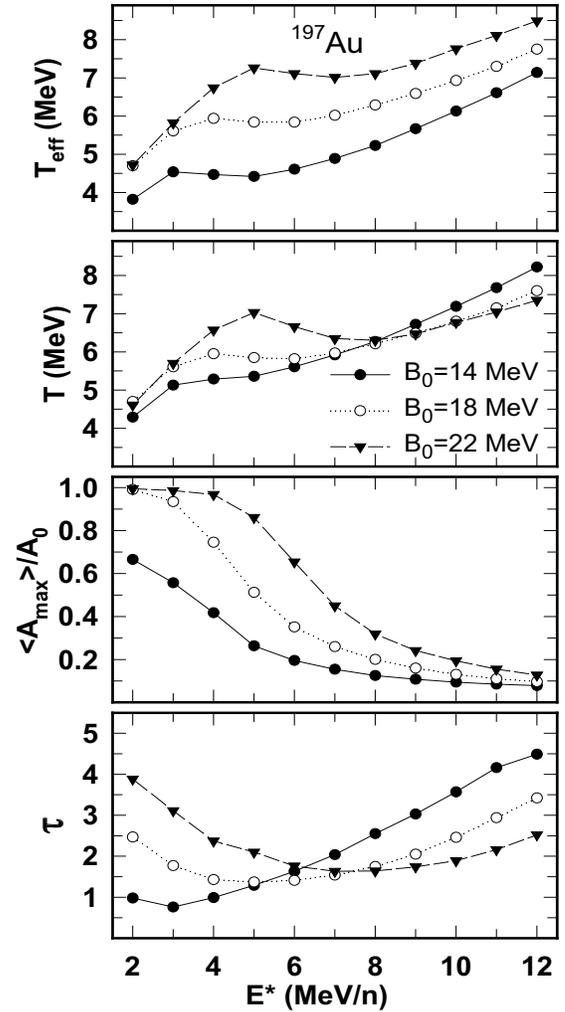}
\caption{\small{SMM calculations of characteristics of hot
fragments 
from Au sources for different
surface energy coefficients $B_0=14, 18, 22$~MeV as a function of the
excitation energy $E^*$. Top two panels: effective temperature $T_{\rm eff}$,
and freeze-out temperature (caloric curves);
middle panel: reduced mass number of the maximum fragment
$A_{\rm max}/A_0$; bottom panel: $\tau$ parameter. }}
\end{figure}

The change in the disintegration process can be conveniently characterized 
by $\tau_{\rm min}$, i.e. the minimum value which $\tau$ assumes as a 
function of the excitation energy. The region near $\tau_{\rm min}$ 
corresponds 
to the transition from channels with one big residual fragment to
multifragmentation into several small fragments. In the nuclear matter
this transition is usually associated with the coexistence region 
of the liquid-gas phase transition. Both, $\tau_{\rm min}$ and the 
corresponding 
effective temperature $T_{\rm min}$, vary as a function of the surface 
energy, as
demonstrated in Fig.~6. As one can see, the three considered sources with 
different isospin content exhibit similar trends but the $\tau_{\rm min}$ 
for the neutron-rich sources is higher.

\begin{figure} [tbh]
\includegraphics[width=8cm,height=11cm]{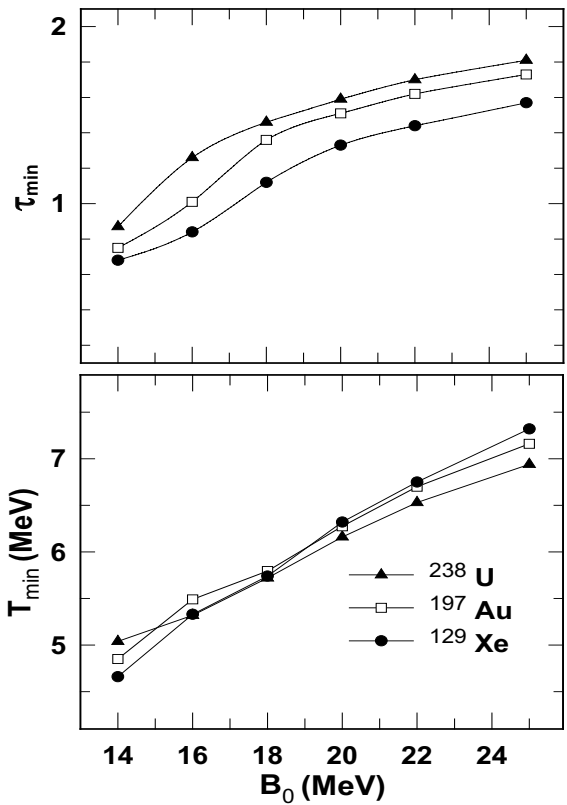}
\caption{\small{Minimum value $\tau_{\rm min}$ of the $\tau$ parameter
for hot fragments and the corresponding effective 
temperature $T_{\rm min}$, extracted from SMM calculations for sources
with different neutron richness $^{238}$U, $^{197}$Au, and
$^{129}$Xe, as a function of the surface energy coefficient $B_0$.
}}
\end{figure}

\section{Experimental data on the $\tau$-parametrization of the charge 
distributions}
\label{exp}

In order to extract information about the surface energy in
multifragmentation we compare the calculations with
experimental data. The ALADIN collaboration has performed exclusive 
measurements of multifragmentation events for projectile-like sources with 
different isospins \cite{ALADIN} and has made these data available. 
Measurements have been made with $^{238}$U, $^{197}$Au, and $^{129}$Xe 
projectiles
at the laboratory energy of 600 MeV per nucleon, 
interacting with different targets ranging from Be to U. 
The power law parameters $\tau$, obtained from fits to the charge distributions
in the range $5 \leq Z \leq 15$, are shown in Fig.~7 as a function of the 
bound charge $Z_{\rm bound}$ (the total charge accumulated in fragments with 
$Z\geq2$) divided by the projectile charge $Z_0$. 
Small $Z_{\rm bound}$ values correspond to high excitation energies
of the sources which disintegrate predominantly into very
light clusters ("fall" of multifragmentation). Large values of
$Z_{\rm bound}$ correspond to low excitation energies, at which the
decay changes its character from evaporation/fission
like processes to multifragmentation ("rise" of multifragmentation). 

The chosen target has no influence on the extracted $\tau$ parameter, as
demonstrated in the bottom part of Fig.~7 for the case of the Au and
U projectiles. This is consistent with the target invariance generally
observed for the charge correlations of projectile fragmentation
\cite{ALADIN}. These invariance properties extend also to
projectiles with different $Z_0$ if the data are scaled accordingly.
For example, the IMF multiplicities divided by $Z_0$ (the reduced 
multiplicities) are practically identical if plotted as a 
function of the reduced bound charge $Z_{\rm bound}/Z_0$. This has been 
shown in \cite{ALADIN} for the case of U, Au and Xe projectiles 
with the same data sets as used here. 
The same holds for the $\tau$ parameters which, however, exhibit 
some variations 
at small and large $Z_{\rm bound}$ (Fig.~7, top). 
Differences at small $Z_{\rm bound}$ may occur for several reasons. 
The uncertainty
of determining $Z_{\rm bound}$ for highly fragmented partitions,
arising from the finite acceptance of the spectrometer for He and Li 
fragments \cite{ALADIN}, may be slightly system dependent which, 
in principle, could be taken into account by using an experimental filter. 
Restricting $Z_{\rm bound}$ to small values will, furthermore, generate 
constraints for the accepted partitions 
which affect the resulting $Z$ distributions \cite{kreutz}. 
The fitting within a fixed $Z$ interval while scaling $Z_{\rm bound}$ could 
thus possibly lead to systematic deviations in $\tau$ for different 
projectiles which, however, are small as evident from the data. 

At large $Z_{\rm bound}$, the observed differences are slightly larger and 
more 
systematic, and no experimental constraints of this type exist. The detection 
efficiency of the ALADIN spectrometer is, furthermore, very high for
$Z \geq 5$, essentially 4$\pi$ in the projectile frame, and the experimental
trigger was adapted for the fragmentation channels \cite{ALADIN}.
In the following, we will, therefore, concentrate on the analysis of the
observed differences in the behavior of $\tau$ for sources with different
isospin at the "rise" of multifragmentation ($Z_{\rm bound}/Z_0 \geq 0.5$). 
Because
of the large acceptance of the detector, the  theoretical predictions
will be directly compared to the data without the use of 
an experimental filter.

\begin{figure}
\includegraphics[width=8cm,height=10cm]{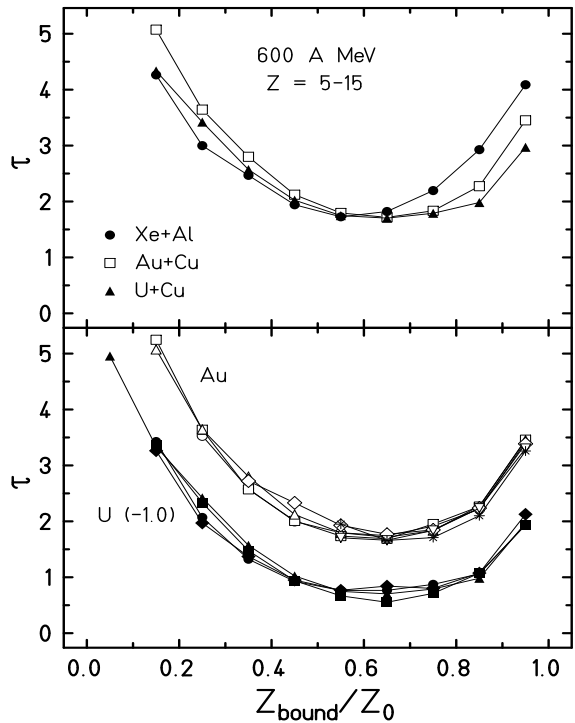}
\caption{\small{Power law parameters $\tau$ obtained from fitting
the fragment charge distributions obtained in ALADIN experiments
in the range $5 \leq Z \leq 15$ as a function of the reduced bound charge
$Z_{\rm bound}/Z_0$ for the fragmentation of $^{129}$Xe, $^{197}$Au and 
$^{238}$U
projectiles of 600 MeV/nucleon. Top panel: three systems with
different projectiles as indicated; bottom panel: results for collisions
of $^{197}$Au with Be, C, Al, Cu, In, Au targets and of $^{238}$U with
Cu, In, Au, U targets. }}
\end{figure}

\section{Theoretical calculations with ensembles of equilibrated sources}
\label{theo}

In the most general consideration the fragmentation process can be
subdivided into several stages as follows. (1) A dynamical stage
leading to the formation of equilibrated nuclear systems, (2)
disassembly of the system into individual primary fragments, and (3)
deexcitation of hot primary fragments. Even though the first
stage may be described by a dynamical model, such as the
intranuclear cascade model, it is more practical to determine
an ensemble of equilibrated sources, produced after the nonequilibrium 
stage, by
direct comparison with experimental data \cite{Botvina92}. In the
following, we consider the ensembles of equilibrated sources extracted
in refs.~\cite{Botvina95,Xi} for the Au projectiles, which
provide a very good description of the ALADIN data. We use also
the same version of the SMM code described fully in \cite{SMM}.
As was found in ref. \cite{Botvina95}, the average masses of equilibrated 
sources $A_s$ may be parametrized as 
$A_s/A_0 =1-a_1(E^*/A_s)-a_2(E^*/A_s)^2$, where 
$E^*$ is the excitation energy of the sources in MeV, and $A_0$ is the 
projectile mass. Here is $a_1$=0.001 MeV$^{-1}$, and 
$a_2\approx$0.009--0.015 MeV$^{-2}$ is slightly depending on the projectile 
energy. It should be noted that the adopted correlation of the mass number 
with the excitation energy is consistent with dynamical simulations 
\cite{Konopka,Barz}, as well as with results of other statistical 
models \cite{Botvina95,Gross,Raduta}. 
Besides providing a very good description of 
fragment production at different $Z_{\rm bound}$, the calculations with this 
kind of ensemble reproduce also correctly the behavior of the caloric 
curve \cite{Xi}. 

Our present analysis is moderately sensitive to the relation between $A_s$ 
and $E^*$ because we extract $\tau$ in different $Z_{\rm bound}/Z_0$ bins, 
which are 
correlated with the excitation energy (per nucleon) of the source. 
Possible variations of the relative weights of the sources in the 
ensembles have practically no effect on the $\tau$  observable, as 
was confirmed by performing calculations with different weights. 
The typical relative yields of sources used for the case of 
$^{197}$Au projectiles are shown in Fig.~8 as a function of their mass 
and excitation energy. The parameter $a_2$=0.012 was used in this case. 
The same relation between masses and excitation energies is also taken 
for the other projectiles. 
It is assumed that the $N/Z$ ratio of all sources in 
the ensemble is the same as in the projectiles. The dynamical
calculations show that this is a quite reasonable assumption
for sources with moderate excitation energies. A very small decrease 
of the $N/Z$ ratio is predicted only for the sources with very high 
excitation energy \cite{Botvina02,LeFevre}. This should not qualitatively 
influence the present analysis in the region of the multifragmentation rise. 

\begin{figure}
\includegraphics[width=8cm,height=8cm]{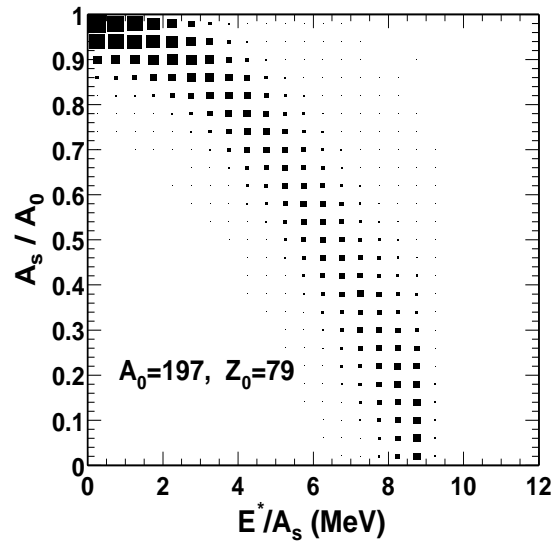}
\caption{\small{Ensemble of hot thermal sources represented in
a scatter plot of reduced mass number $A_s/A_0$ versus excitation energy
$E^*/A_s$ for the fragmentation of $^{197}$Au projectiles, as used in the SMM
calculations. The intensity of the individual sources is proportional
to the area of the squares.  }}
\end{figure}

The SMM calculations with these ensembles for different projectiles 
predict a scaling of the fragment multiplicities, as well as other 
observables, in agreement with the experimental data \cite{ALADIN}. 
This is illustrated in Fig.~9, where we plot the reduced IMF 
multiplicities versus $Z_{\rm bound}/Z_0$. In order to demonstrate the 
sensitivity 
of this correlation to the ensemble parameters and to the surface 
energy, we compare calculations obtained with different parameter sets. 
For $^{197}$Au we show results for the ensemble parameters $a_2$=0.011 
(full squares) and $a_2$=0.012 (empty squares), obtained with the surface 
term $B_0$=18 MeV. For $^{238}$U we compare the cases $B_0$=17 MeV 
(empty triangles) 
and $B_0$=18 MeV (full triangles) for ensembles generated with $a_2$=0.011
and, for the ensembles representing the fragmentation of $^{129}$Xe, generated
with $a_2$=0.012, we show the results for $B_0$=19 MeV 
(full circles) and $B_0$=20 MeV (empty circles). 
In this way we cover the range of parameters 
representative for the following analysis of the data. 
One can clearly see a very good scaling behaviour, similar to the 
experimental one demonstrated in Fig.~9 of ref. \cite{ALADIN}. 
A good description of the IMF multiplicities allows us to adjust the ensemble 
parameters and start the $\tau$ analysis.  
This analysis allows us to extract more detailed information on the
multifragmentation process, since the $\tau$ fit 
in the range $Z$=5--15 represents a new sensitive observable. 
It is important to mention that the maximum IMF production is reached 
at $E^{*}$ around 5--6 MeV/nucleon, when the sizes of the sources are 
relatively large (Fig.~8). 

\begin{figure}
\includegraphics[width=8cm,height=8cm]{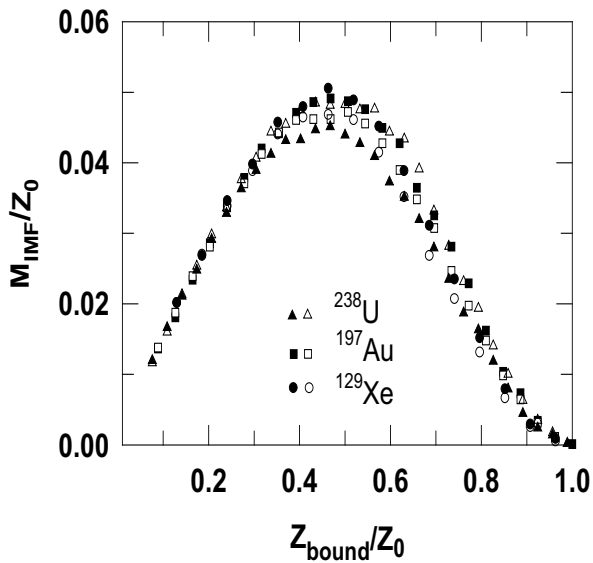}
\caption{\small{Mean IMF multiplicities versus $Z_{\rm bound}/Z_0$,
the bound charge divided by the projectile charge, predicted by the 
SMM ensemble calculations for three different projectiles. 
The meaning of the symbols is explained in the text. 
}}
\end{figure}

The results of the SMM calculations for ensembles of the sources with 
$a_2$=0.012 produced with Xe, Au, and U projectiles are shown in Fig.~10.
In the case of Au, the obtained $\tau$ parameters are close to the experimental
values. This remains also valid in the region of the multifragmentation 
"fall", i.e. at high excitation energies and small $Z_{\rm bound}$, even 
though 
the sources become small and have very large size variations (Fig.~8).
However, in the region of the multifragmentation "rise" the 
theory gives larger $\tau$ for the neutron--rich U projectile
than for the neutron--poor Xe projectile. This is opposite to the
experimental observation. Below we investigate possible modifications 
of the SMM which can explain this discrepancy. 

\begin{figure}
\includegraphics[width=8cm,height=8cm]{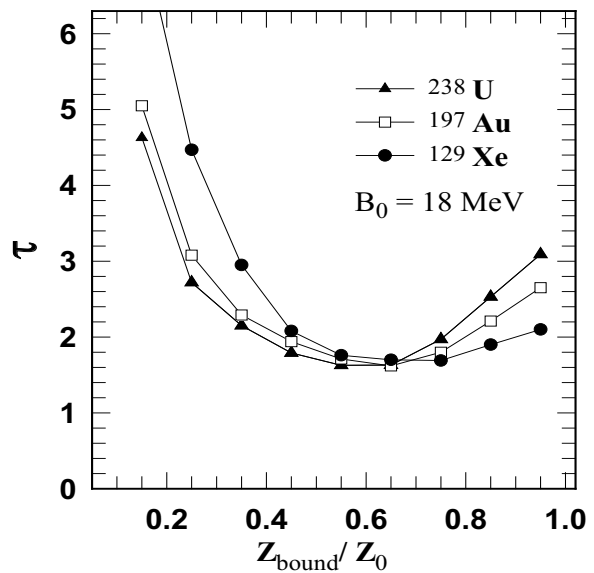}
\caption{\small{SMM ensemble calculations of $\tau$
parameters for cold fragments versus $Z_{\rm bound}/Z_0$ for
$^{238}$U,  $^{197}$Au, and $^{129}$Xe projectiles. The surface
energy coefficient is $B_0$=18 MeV for all sources. }}
\end{figure}

First, we have tested whether one can improve the agreement with the 
experiment by simply using mass formulae with 
an explicit isospin dependence of the surface energy. 
We have implemented the Myers-Swiatecki 
mass formula \cite{Myers} in the SMM, instead of the standard liquid-drop 
description of hot fragments. In this case the surface energy coefficient is 
written as $B_0=18.56 (1-1.79(1-2Z/A)^2)$ MeV. 
The other parameters were also modified according to ref. \cite{Myers}. 
The results of the SMM calculations with the Myers-Swiatecki parametrization 
are shown in Fig.~11 for the ensembles with $a_2$=0.012, 
for all three projectiles. 
The resulting values of $\tau$ for different projectiles 
at large $Z_{\rm bound}$ are less spread than the results with constant 
$B_0 = 18$~MeV (Fig.~10) but they do not fit the experimental trend. 
Moreover, the general agreement 
with the experiment becomes even worse, since the minimum of $\tau$ is
considerably shifted to larger values of $Z_{\rm bound}$. Also 
$\tau_{\rm min}$ is less than 1.5,
which is too low in comparison with the experimental value of about 1.7. 
Calculations with other ensembles did not improve the agreement with 
the experiment. This suggests that a more detailed investigation of the 
liquid-drop parameters of fragments in multifragmentation reactions is 
needed. 

\begin{figure}
\includegraphics[width=8cm,height=8cm]{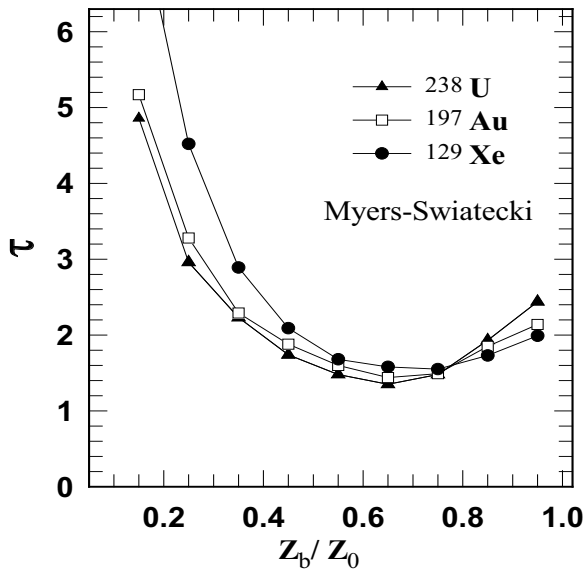}
\caption{\small{SMM ensemble calculations of $\tau$
parameters for cold fragments versus $Z_{\rm bound}/Z_0$ for
$^{238}$U,  $^{197}$Au, and $^{129}$Xe projectiles. The 
Myers-Swiatecki mass formula was used for the hot primary fragments. }}
\end{figure}

Returning to the standard SMM, we have performed calculations for the 
same projectile ensembles but with other surface coefficients $B_0$. 
We have selected four intervals $Z_{\rm bound}/Z_0$=0.8--0.9,
0.7--0.8, 0.6--0.7, and 0.5--0.6, characterizing the "rise" 
of multifragmentation with excitation energy, 
for a detailed comparison with the experiment.
As one can expect from Fig.~5, in order to fit the higher $\tau$ observed for
neutron-poor Xe sources at low excitation energies, it is necessary
to increase the surface coefficient $B_0$ for the corresponding
hot fragments. As known from previous analyses (see, e.g., \cite{ogul}),
the secondary decay processes should not significantly influence $\tau$ in
this region. For the secondary deexcitation calculations we have used the
standard evaporation and Fermi break-up models implemented in the
SMM \cite{Botvina87}. The same deexcitation code was used in refs. 
\cite{Botvina95,Xi} for describing the ALADIN data.

The theoretical evolution of $\tau$ with $B_0$ in different
$Z_{\rm bound}$ bins is shown in Fig.~12.
In order to achieve agreement with the experimental
data, $B_0$ should change with excitation energy when going
from compound-like processes to full multifragmentation. In the bin of the
largest $Z_{\rm bound}/Z_0=0.85$, this requires $B_0\approx20.0$, 18.1, and
17.1 MeV for the Xe, Au, and U ensemble sources, respectively.
At $Z_{\rm bound}/Z_0$=0.65 and 0.55, however, 
in the region of minimum $\tau$ and
full multifragmentation, the values of the surface tension are
around $B_0\approx$18--19 MeV for all three sources without essential 
variation.

\begin{figure}
\includegraphics[width=8cm,height=14cm]{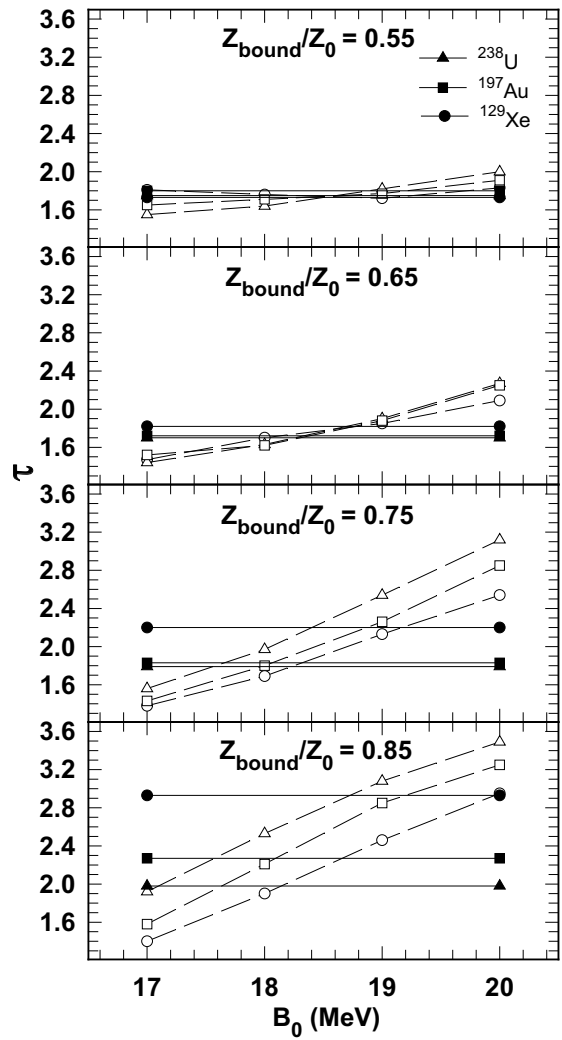}
\caption{\small{Comparison of the $\tau$ parameters obtained from
the SMM ensemble calculations as a function of $B_0$
(dashed lines and open symbols) with the experimental results
(solid horizontal lines and full symbols) for U, Au, and Xe projectiles, 
in different $Z_{\rm bound}/Z_0$ bins.
 }}
\end{figure}

We have also analyzed other ensembles and combinations of ensembles for 
different projectiles, which can reproduce the IMF scaling shown 
in Fig.~9 and in ref. \cite{ALADIN}. In addition, we require that these 
ensembles reproduce the experimental $\tau$ in the whole range of 
$Z_{\rm bound}$ by varying $B_0$. 
We have found that such ensembles should have parameters in between 
$a_2$=0.011 and $a_2$=0.013. 
For control, we have also tested the extreme cases as, e.g., 
$a_2$=0.011 for the ensembles of U sources and $a_2$=0.013 for the 
Xe sources. Even for these ensembles, we have obtained a similar evolution of 
the $B_0$ isospin dependence as a function of the excitation energy: 
while $B_0$ depends strongly on $N/Z$ at large $Z_{\rm bound}$, it becomes 
nearly independent of isospin in the region of full multifragmentation. 

Additional calculations were performed with another secondary
decay code developed in ref. \cite{nihal}. This code starts
with the modified surface tension for hot fragments in the freeze-out
volume, and then changes the fragment masses during the evaporation process 
in such a way that the normal
surface tension is restored for cold fragments. It was found that
the final results may deviate from the standard calculations by not
more than a few percent, and that all trends regarding $B_0$ as a function
of the neutron richness of fragments remain the same.

\section{Discussion of the results}
\label{disc}

As was postulated earlier (Sect.~\ref{surf}), 
the surface term is a function of two parameters, $B_0$ and the critical
temperature $T_{\rm c}$. Therefore, 
a dependence of $T_{\rm c}$ on the neutron content could be an
alternative explanation of the isospin dependence of the surface
tension in different domains of excitation energy. This possibility was 
investigated by performing an analysis with a possible $T_{\rm c}$ evolution,
as suggested in refs. \cite{ogul,karna}.
It was found that, in order to increase $\tau$ to the experimental value
observed for Xe sources at
large $Z_{\rm bound}$, $T_{\rm c}$ would have to be increased to 
$T_{\rm c}\approx24$~MeV. In the same
$Z_{\rm bound}$ range, in order to reproduce the value for U sources,
$T_{\rm c}$ would have to be decreased to $T_{\rm c}\approx14$~MeV. 
A $T_{\rm c}$ variation of this
magnitude in a narrow $N/Z$ range (from 1.39 to 1.59) is not supported by
theoretical calculations \cite{Ravenhall}, which predict a variation
of $T_{\rm c}$ within 2--3\% only. In addition, large $T_{\rm c}$ 
differences for the 
projectile sources will result in a significant violation of the scaling
behaviour of fragmentation at high excitation energies, in contradiction to
the experiment.

We have mentioned that for small sources, at $Z_{\rm bound}/Z_0 <$0.5, the 
experimental filter should be applied for a precise determination of 
$Z_{\rm bound}$ in simulated events. This would be important for our method 
since at high excitation energy, as one can see from Figs.~4 and 10, the 
$\tau$ extracted for the sources of different isospin nearly coincide, while 
the $\tau$ values after the ensemble calculations are different, when 
plotted versus the reduced bound charge. 
This is a consequence of dealing with the small sources and 
large fluctuations at the multifragmentation 'fall'. 
Also, in the region of small $Z_{\rm bound}$, after SMM calculations, 
many observables are less sensitive to differences in the ensembles of 
sources: As evident from Fig.~9, the slope of $M_{IMF}/Z_0$ versus 
$Z_{\rm bound}/Z_0$ at small $Z_{\rm bound}$ 
is rather universal for all ensembles and values of $B_0$, 
and it coincides with the data \cite{ALADIN}. 
Without independent constraints on the sources in this region, 
it is not very reliable to extract fragment properties by only 
using the $\tau$ parametrization of the charge yields. 
Probably, involving isotope distributions of produced 
fragments and measuring neutrons \cite{sfienti,trautmann} will help to 
determine the ensembles. We do not touch this problem here but just 
note that extraction of $B_0$ as a function of the neutron richness at 
small $Z_{\rm bound}$ will depend essentially on the excitation energies of 
the sources. As one can see from Fig.~5 (bottom panel), if the average 
excitation energy is around 7--8 MeV/nucleon, the sensitivity of $\tau$ 
to the surface coefficient is lost at $B_0 \approx 18-22$ MeV. 

Besides their different $N/Z$ ratios, the three projectiles differ also
in their mass. To first order, this is considered by performing the analysis
as a function of scaled variables but a residual mass effect may possibly
exist. 
The calculations presented in ref.~\cite{nihal} show, however, that 
for relatively large sources $A_s \sim 100$ a
difference in the $N/Z$ ratio has a larger effect on $\tau$ than a difference
in the mass. Moreover, calculations for the fragmentation
of nuclei with different
$N/Z$ and the same mass predict differences in $\tau$ of the same magnitude
as obtained here for the considered Xe, Au and U nuclei (Fig.~10). This
justifies our conclusion that the deviation of the standard SMM calculations 
from the experimental data at low excitation energies is due to the 
isospin dependence of the surface coefficient. This conclusion can 
be tested with fragmentation data in experiments with isotopically 
different isobars.

\begin{figure}

\includegraphics[width=8cm,height=8cm]{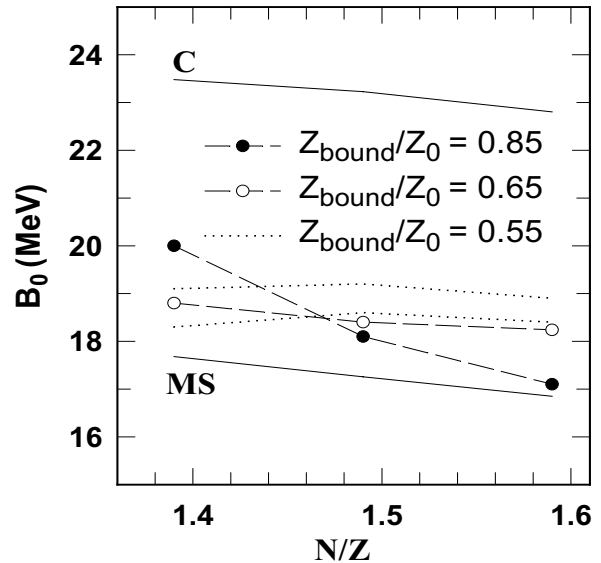}
\caption{\small{The extracted $B_0$ for ensemble sources with
different $N/Z$ ratios at the onset ($Z_{\rm bound}/Z_0$=0.85 bin), and at
the region of full multifragmentation ($Z_{\rm bound}/Z_0$=0.65 and 0.55 
bins). The two dotted lines for the lowest bin indicate upper and 
lower limits given by our method. 
The $B_0$ obtained from Cameron (C) and Myers-Swiatecki (MS) mass 
formulae for cold nuclei are shown by solid lines for
illustration. }}
\end{figure}

It is instructive to compare the surface coefficients extracted above 
with the ones used usually for 
cold isolated nuclei. In Fig.~13 we show the behavior of this 
coefficient versus the $N/Z$ ratio as extracted from the 
Cameron \cite{Cameron} and Myers-Swiatecki \cite{Myers} mass formulae 
by retaining terms proportional to $A^{2/3}$. The considered $N/Z$ range
corresponds to the neutron-richness of the sources under investigation.
Both mass formulae predict a decrease of $B_0$ with increasing
$N/Z$ which is necessary in order to reach agreement with experimental
mass tables. In multifragmentation,
as was shown in refs. \cite{Botvina02,Shetty05,Botvina01}, the neutron
content of hot fragments is proportional to the neutron richness of the
sources, since most neutrons are accumulated in the fragments. The
values of $B_0$ extracted from the experiment exhibit
the same decreasing trend with $N/Z$ at low excitation energy
and they remain within the $B_0$ range predicted by the mass formulae.
Thus, at the rise of multifragmentation, when a heavy residue 
and few small fragments are formed, $B_0$
resembles the values used in phenomenological parameterizations for 
isolated nuclei. On the other hand, at full multifragmentation,
when the system breaks up into intermediate mass fragments,
the surface energy coefficient becomes nearly independent of the
neutron richness. 
This is particularly evident for the bins $Z_{\rm bound}/Z_0$=0.65, where 
the decrease in $B_0$ with $N/Z$ is very small, and 
$Z_{\rm bound}/Z_0$=0.55, at which the $B_0$ consistent with $\tau$ falls 
between the two dotted lines (Fig.~13). These upper and lower 
limits correspond
to the 2\% statistical uncertainty in the numerical determination of $\tau$, 
typical for both, the SMM simulations and the experimental data. 

These observations indicate that nuclear fragments are brought to a new
physical environment. Their properties change in the hot medium consisting 
of nucleons and other fragments at subnuclear freeze-out densities, 
around $ 0.1 \rho_0$. 
Also, modifications of the symmetry energy have been observed in these 
reactions \cite{LeFevre}. Here, 
we do not suggest a comprehensive theoretical explanation of these 
effects, but rather point at the new physical conditions reached in
multifragmentation reactions in comparison with isolated nuclei. The
origin of these effects could be a residual nuclear interaction between
clusters, in particular, the exchange of neutrons between nuclear 
fragments. The role of the Coulomb forces is also important. 
The Coulomb interaction between clusters
leads to a modification of their average charges \cite{Botvina87}.
Other consequences of the Coulomb interaction, like a shift of the
proton distributions with respect to the neutron distributions in hot 
fragments \cite{jandel}, may additionally contribute to this effect. 

Finally we point to the uncertainty connected with
the lacking knowledge of the $N/Z$ ratio of the produced equilibrated sources
after the initial dynamical stage. The present dynamical
codes, like the intranuclear cascade (INC) \cite{Botvina02}, and the
relativistic quantum molecular dynamics (RQMD) \cite{LeFevre,Gaitanos},
predict a very small, less than 5$\%$, change in comparison with the neutron
richness of the projectile nucleus, since interactions of neutrons and
protons are very similar at relativistic energies. This change is
expected to be small for all projectiles, therefore, the difference
in neutron content should be
preserved for the ensembles of projectile sources.
Similar conclusions can also be drawn on the basis of existing
experimental data:
As was established in recent ALADIN experiments, fragments coming
from neutron-rich projectiles remain essentially neutron-rich
as compared with fragments from neutron-poor projectiles \cite{sfienti}.
At the same time, the number of produced spectator neutrons is correlated 
directly with the neutron richness of the projectiles \cite{trautmann}. 
These results indicate that the initial difference in $N/Z$ ratios for
different projectiles is not washed out for the ensemble of
sources during the nonequilibrium stage. 

\section{Conclusions}
\label{concl}

In this paper we have demonstrated that the surface properties of
hot fragments can be studied in multifragmentation reactions. 
Within the Statistical Multifragmentation Model we have shown how 
modifications of the fragment surface energy influence the 
fragment production and thermodynamical characteristics of the 
system at the freeze-out volume. 
By comparing with the ALADIN data we have found that 
in the regime of the ''rise'' of multifragmentation 
the observed isospin dependence of the power-law parameter $\tau$ of fragment
charge distributions cannot be described by the SMM 
if the standard surface-term coefficient is used. The data can be 
reproduced if a moderate dependence on the $N/Z$ ratio of the produced
fragments is introduced in the surface term. At low excitation energies 
this contribution resembles that 
obtained with advanced mass formulae for ground-state 
nuclei. At higher excitation energies, this $N/Z$ dependence gradually 
disappears. 

The observed modification of properties of hot nuclei in comparison
with cold (or slightly excited) isolated nuclei are possibly 
caused by the hot environment in which the nuclei are imbedded, and
where they can interact with nucleons and other fragments. The resulting
in-medium modifications can go beyond the disappearance of shell effects
and affect also the main liquid-drop parameters.
Our conclusions are substantiated by a very good agreement
with experimental data demonstrated by the SMM. In the future, fully
microscopic calculations of the effects of the interacting medium
would be very desirable, however, many problems still remain in
constructing such a many-body theory.

More generally, we conclude that multifragmentation reactions offer
a new possibility for investigating nuclei under conditions essentially
different from those accessible in nuclear structure studies at
low energies. Besides the
obvious application of the results for better describing
nuclear reactions at intermediate and high energies, this has
far-reaching consequences for astrophysics. In many astrophysical
situations, hot matter at subnuclear densities is clusterized
into nuclei as, e.g., in neutron star crusts and supernovae.
The properties of these nuclei
are crucial for both microscopic (weak interactions) and
macroscopic (collapse dynamics and shock propagation) processes
in stellar matter. New and important information in this direction
may be expected from the analysis of future experimental data on 
multifragmentation. 

\begin{acknowledgments}

The authors would like to thank the ALADIN Collaboration for the
permission to use the charge-correlation data for the present
analysis. A.B. thanks FIAS Frankfurt and GSI Darmstadt
for support and hospitality, R.O. thanks TUBITAK-DFG cooperation, 
and N.B. thanks Selcuk
University-BAP(05401052) and the GSI Darmstadt for financial support. 
This work was partly supported by the grants RFBR 05-02-04013, 
and NS-8756.2006.2 (Russia). 

\end{acknowledgments}

\end{document}